# Experimental Study on Boiling of Nanofluids in Copper Foam

Kai-Xin Hu[a,b,1], Jing-Han Pan[a,b]

[a]Zhejiang Provincial Engineering Research Center for the Safety of Pressure Vessel and Pipeline, Ningbo University, Ningbo, Zhejiang 315211, China

[b]Key Laboratory of Impact and Safety Engineering (Ningbo University), Ministry of Education, Ningbo, Zhejiang 315211, China

**Abstract**: Nanofluids are suspensions of nanoscale particles (such as metals and their oxides) in base fluids (such as water, oil, or alcohol), which can significantly enhance the heat transfer performance of the base fluid. However, when nanofluids are applied to heat pipes, it is common for nanoparticles to accumulate within the heat pipe's capillary wick, clogging it and increasing thermal resistance. This paper investigates the phenomenon of boiling of water and nanofluids enhanced by copper foam through experimental methods. When the liquid is injected into copper foam placed on a heating plate, some of the liquid is squeezed out along the boundary of the heated surface of the copper foam during boiling. This phenomenon is independent of gravity but related to the hydrophilicity or hydrophobicity of the heating surface. Based on these properties, we design a device to guide the squeezed-out liquid to other locations, offering a promising solution to the problem of nanoparticle accumulation in the heat pipe's capillary wick.

**Keywords**: Nanofluids, Heat pipes, Copper foam, Boiling

## 1. Introduction

Heat pipes are devices that utilize the evaporation and condensation of liquids to transfer heat. They can transport large amounts of heat through small cross-sectional areas without the need for additional components and without consuming energy [1]. Therefore, they have diverse and extensive applications in the field of energy conservation and emission reduction, such as electronic equipment cooling, solar and geothermal energy utilization, and industrial waste heat recovery. As research on heat pipes continues to deepen, improving the heat transfer efficiency has become the main focus of current research.

[1]Corresponding author, Email: hukaixin@nbu.edu.cn

Using nanofluids in heat pipes has been a research hotspot in recent years [2]. Nanofluids [3], as a new type of heat transfer medium formed by adding nanoscale metal or non-metal oxide particles to fluids in a certain manner and proportion, offer enhanced heat transfer capabilities due to their increased surface area. The addition of solid particles makes them more suitable for heat transfer. Additionally, the internal movement of nanoparticles, influenced by forces such as Brownian motion, enhances the energy transfer process between particles and the liquid, thereby increasing thermal conductivity [4].

Another major factor affecting the heat transfer performance of heat pipes is the choice of wick. Foam metal, as a porous material, offers the advantages of a large specific surface area and high thermal conductivity [5-6]. Using it as the wick of a heat pipe can enhance the heat transfer capacity, reduce the tube wall temperature, and provide significant capillary force. In recent years, research on the enhancement of flow heat transfer using copper foam in heat pipes has received more attention [7-8]. However, nanoparticles cannot follow the evaporated gas through the wick of the heat pipe, inevitably leading to their accumulation within the wick, increasing the thermal resistance during evaporation, and easily clogging the wick. Additionally, the role of nanoparticles is primarily limited to enhancing heat conduction in the evaporation zone, and their effect on enhancing convective heat transfer has not been fully utilized. This is the main drawback of nanofluid heat pipes at present.

This paper investigates the phenomenon of boiling in water and nanofluids within copper foam, and based on their boiling characteristics, designs a collection device for splashing liquid during boiling in the copper foam. This device can be used to address the issue of nanoparticle aggregation and clogging within the copper foam.

# 2. Experimental Methods and Results

Figure 1 shows a schematic diagram of our experimental setup. A copper foam (dimensions: $100*50*5mm^3$ , 60ppi, porosity 98%, volume density $0.7g/cm^3$ ) is fixed on a heating plate, which is then heated. Liquid (water and nanofluids) is injected onto the upper surface of the copper foam to be absorbed, and the evaporation and boiling phenomena of the liquid within the copper foam are observed. The z-axis is along the outward normal direction of the heating plate.

2.1 Boiling Experiment for Water

When water boils in the copper foam, bubbles continuously appear at the boundary of the heated surface of the copper foam. This is because the liquid inside the copper foam, during the boiling process, squeezes out a portion of the liquid, which then falls onto the heating plate and continues to boil, producing bubbles.

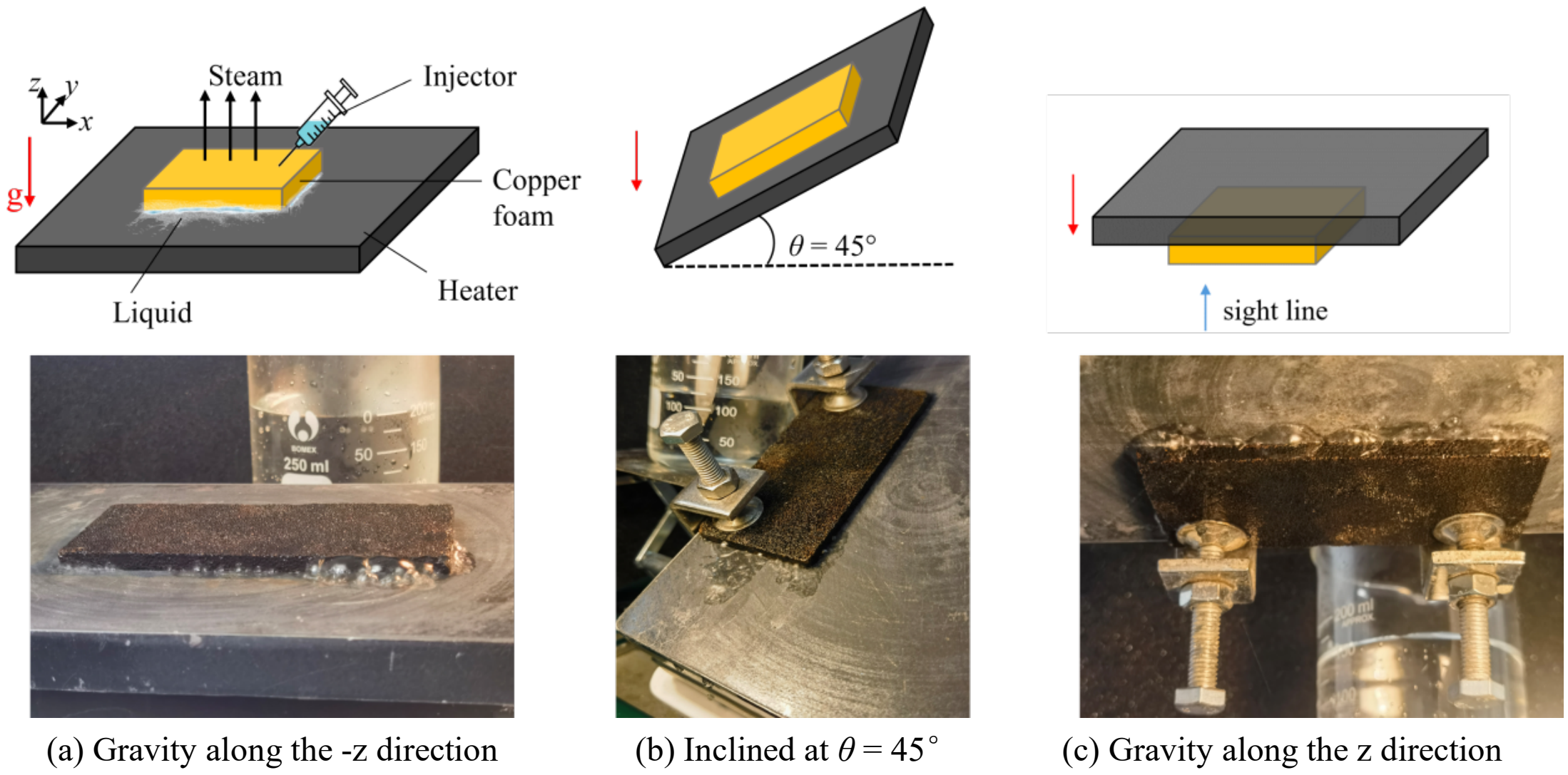

(a) Gravity along the -z direction (b) Inclined at $\theta = 45^\circ$ (c) Gravity along the z direction

Fig. 1 Boiling of Water in Copper Foam

We test three gravity directions in our experiments to study the influence of gravity on this phenomenon. In Figure 1a, the copper foam is placed on the heating plate, which is horizontally positioned; in Figure 1b, the heating plate is inclined at an angle $\theta = 45^\circ$ relative to the horizontal plane; and in Figure 1c, the copper foam is fixed below the heating plate. The experiment found that under all three gravity directions, the boiling liquid would expel bubbles at the edge of the contact surface between the copper foam and the heating plate, proving that the phenomenon is independent of gravity direction. Figure 1c also illustrates that the liquid at the edge is squeezed out due to boiling inside the copper foam, not flowing out under the influence of gravity; otherwise, the experimental phenomenon should be liquid dripping from the lower surface of the copper foam.

To test the influence of the wettability of the heating surface on the experimental phenomenon, we coat the left and right sides of a copper plate with hydrophilic and hydrophobic coatings, respectively (Figure 2a), and then placed this copper plate on the heating stage to serve as the heating surface for the copper foam. During the boiling experiments (Figures 2b and 2c), we observe that bubbles appear on the heating surface coated with the hydrophilic coating, while no bubbles are found on the hydrophobic side. In addition, this phenomenon is also independent of gravity direction. This indicates that we can control the location where the liquid is squeezed out along the boundary by coating the heating surface with hydrophilic/hydrophobic coatings.

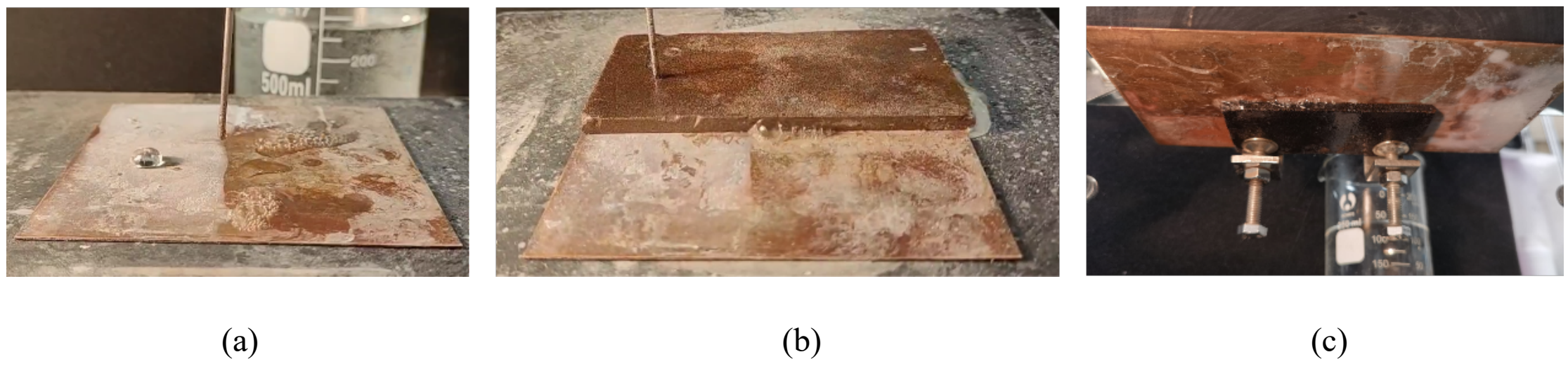

(a) (b) (c)

Fig. 2 Boiling on Hydrophilic and Hydrophobic Surfaces (a) Copper plate coated with hydrophilic/hydrophobic coatings (b) Gravity downward (-z direction) (c) Gravity upward (z direction)

## 2.2 Boiling Experiment for Nanofluids

We drip the nanofluid (CuO, 20wt%, 100nm) onto the heating stage and heat it to boiling. It is observed that during the boiling process, the nanoparticles can not follow the steam into the atmosphere but remain on the heated solid surface, as shown in Figure 3.

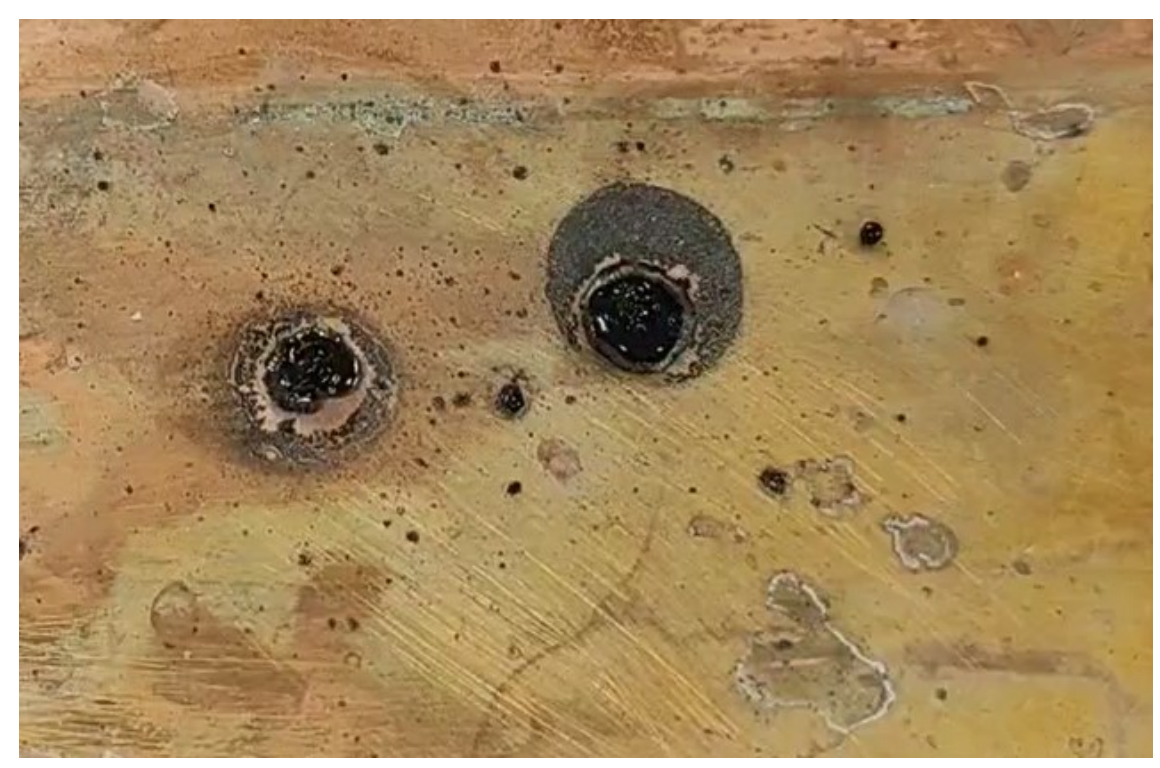

Fig. 3 Nanoparticles left after boiling of nanofluid

We conduct the boiling experiment shown in Figure 1 using the nanofluid in a copper foam. The result show that during the boiling process, the nanofluid also exhibits the phenomenon of liquid being squeezed out of the copper foam, which is also independent of gravity direction, as shown in Figure 4. It can be observed that the black nanofluid is squeezed out from the interior of the copper foam onto the heating plate, and then the nanofluid at the edge of the heated surface of the copper foam continues to boil, with the water turning into steam and the black nanoparticles remaining on the heating plate.

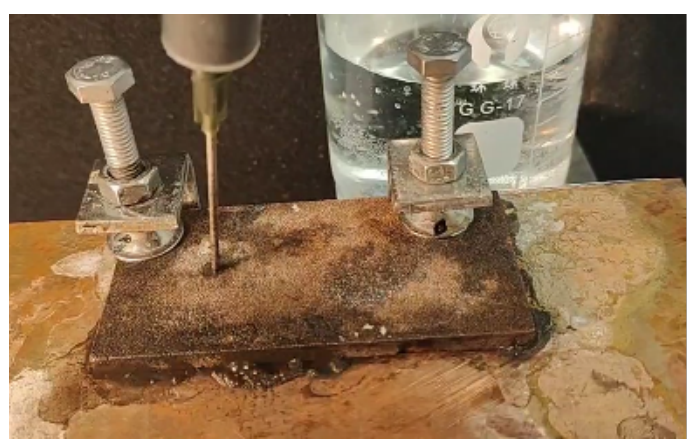

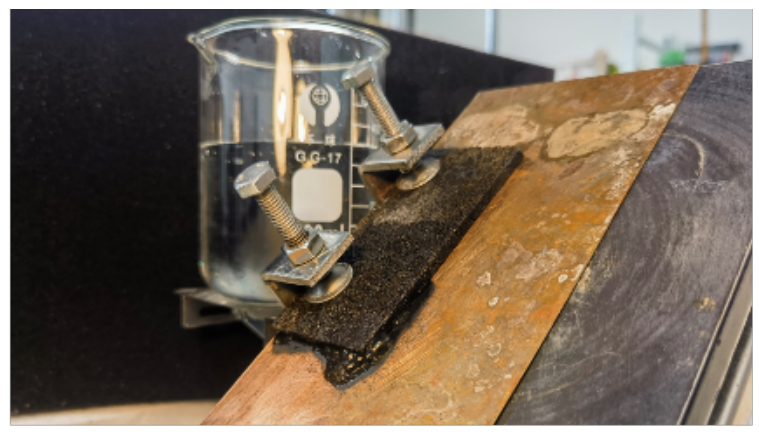

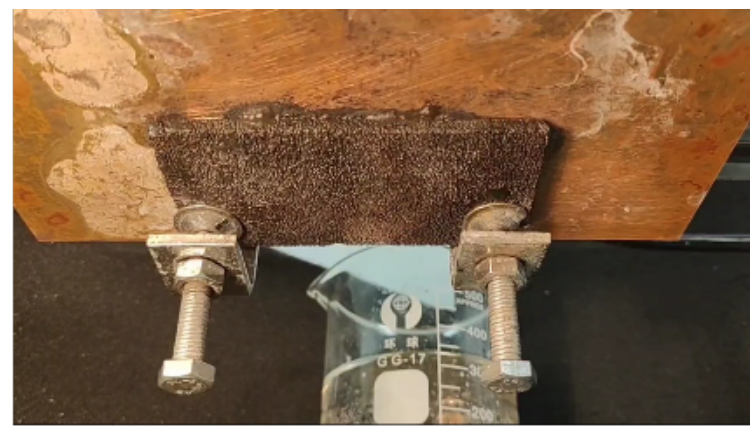

(a) Gravity downward (-z direction) (b) Inclined at $\theta = 45^{\circ}$ (c) Gravity upward (z direction)

Fig. 4 Boiling of Nanofluid in Copper Foam

## 2.3 Diverting the Squeezed Fluid

In Figure 5a, the front part of the copper foam is serrated, with the purpose of increasing the amount of liquid squeezed out from the copper foam during boiling. To direct the squeezed liquid to other locations, we design a liquid collector as shown in Figure 5b, which contains five conical channels. We coat the inner walls of the channels with superhydrophilic coatings to facilitate the suction of the squeezed liquid into the channels.

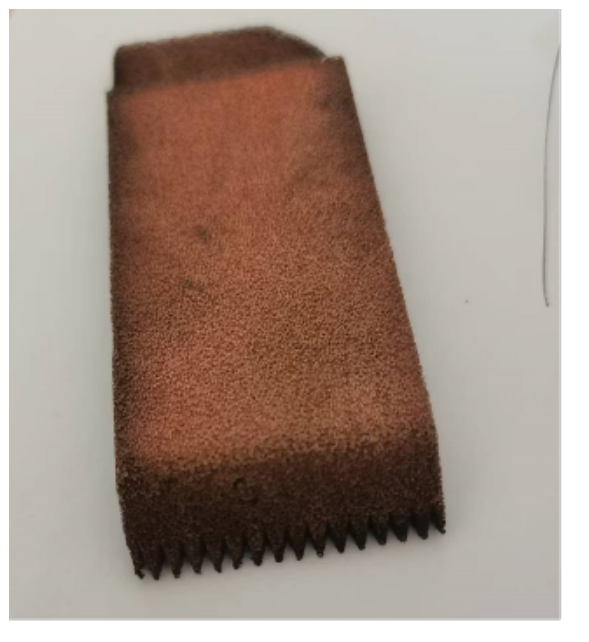

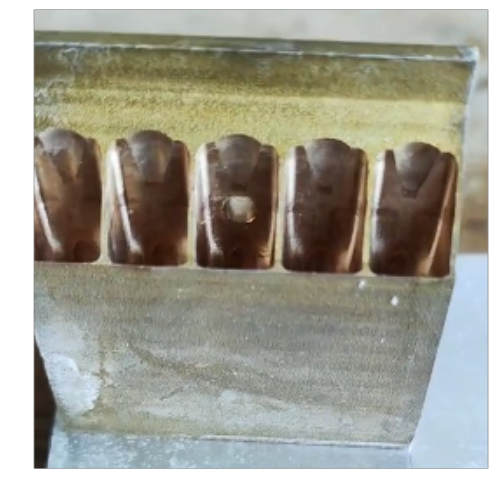

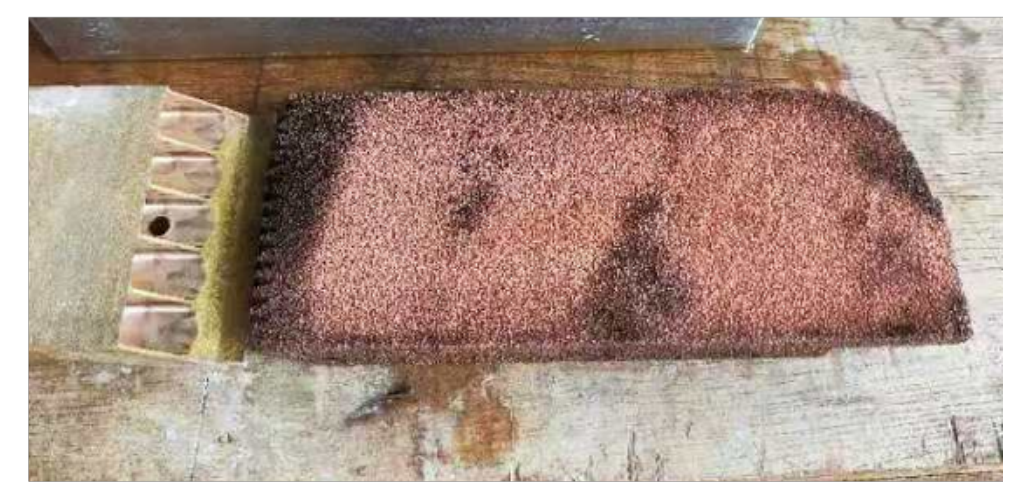

(a) Copper foam (b) Liquid collector (c) Copper foam connected to the liquid collector

Fig. 5 Liquid Collection Device

To control the liquid flow, obstacles are first placed on both sides of the copper foam, as shown in Figure 6a, so that the water during boiling could only splash out from the front. Using the liquid collection device for diversion, as shown in Figure 6b, during the boiling process, the liquid is ejected from the small holes on the liquid collector. If we connect capillary tubes to these holes, the liquid can be drained. By using the designed channels for diversion, we can direct the flow of the liquid squeezed out during boiling.

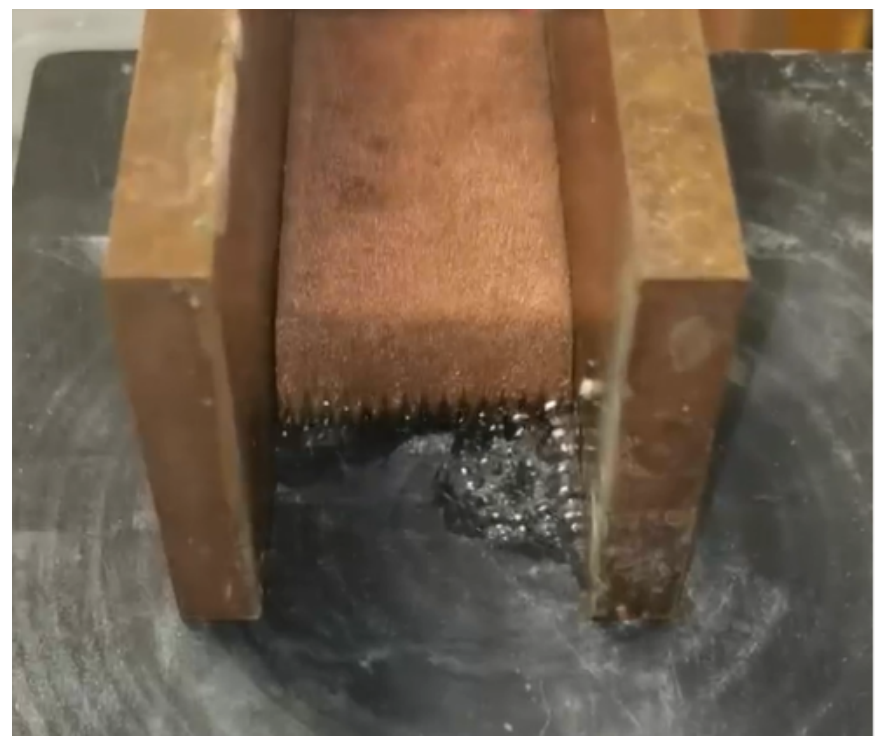

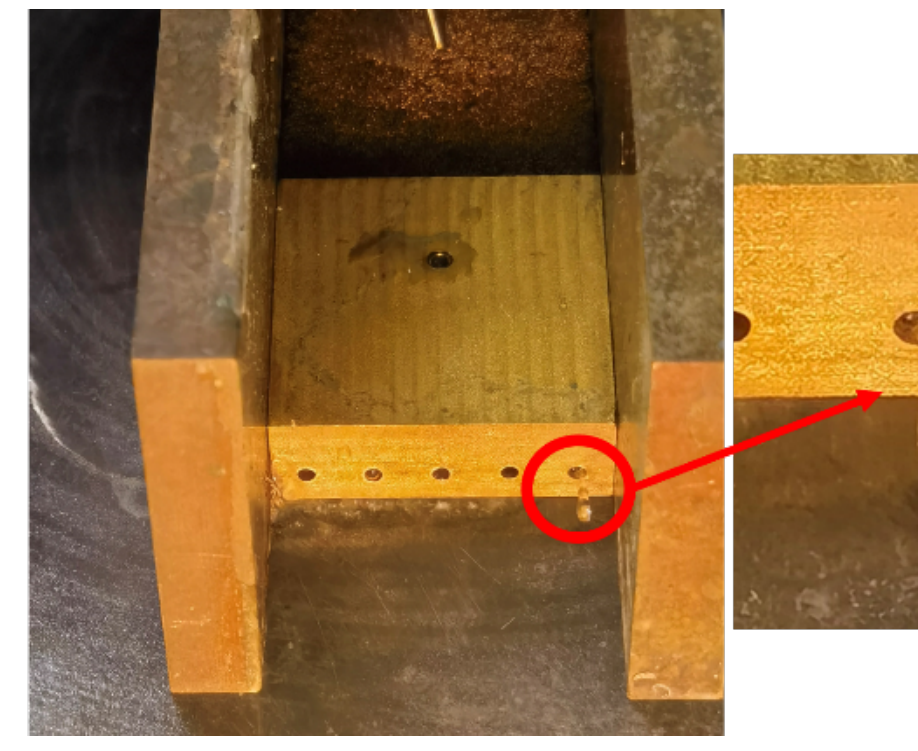

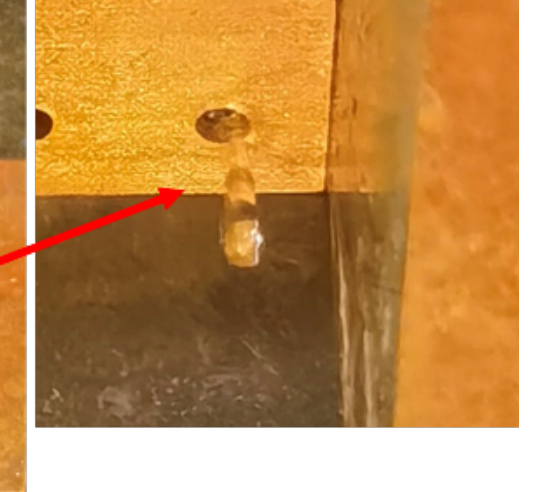

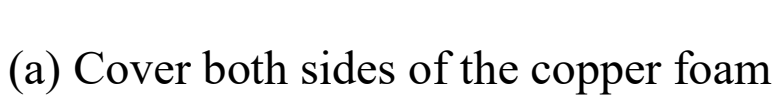

(a) Cover both sides of the copper foam (b) Liquid collection device

Fig. 6 Collecting Boiling Liquid from Copper Foam

Figure 7 illustrates the device for directing the squeezed liquid towards the spiked plate. The system is placed on an aluminum trough, as shown in Figure 7a. When the liquid in the copper foam is heated to boiling by the heating plate at the bottom, part of the liquid is squeezed out and flows towards the liquid collector, then through the capillary to the spiked plate. To enhance boiling, we place a 500g weight on top of the copper foam, ensuring a tighter contact between the copper foam and the bottom aluminum plate. It can be observed that liquid is flowing out at the connection between the capillary tube and the spiked plate. The initially dry spiked surface (Figure 7b) is wetted by the liquid flowing out from the capillary tube during the experiment (Figure 7c), proving that the boiling liquid in the copper foam can achieve directional flow through the device.

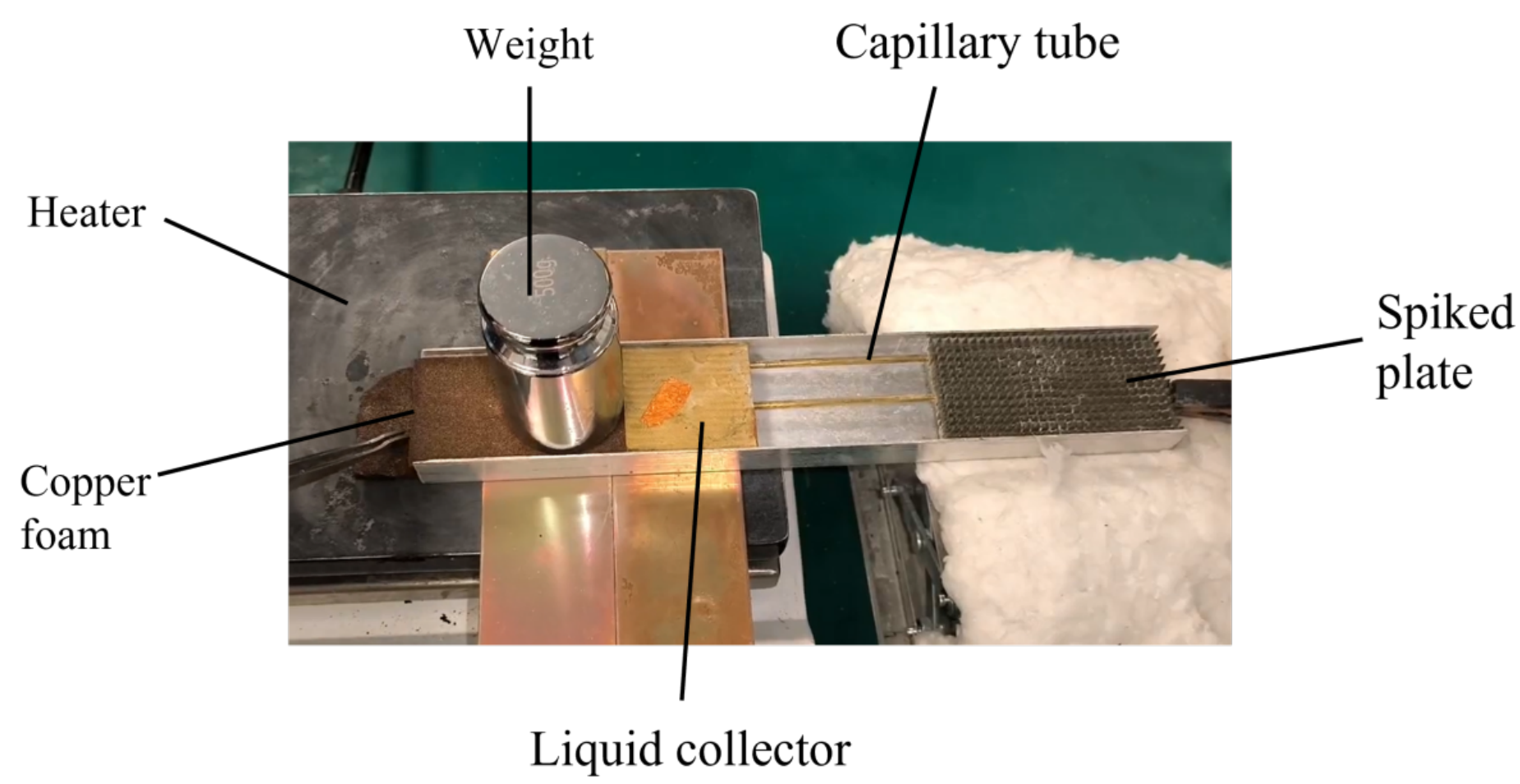

(a) Liquid collection device with drainage

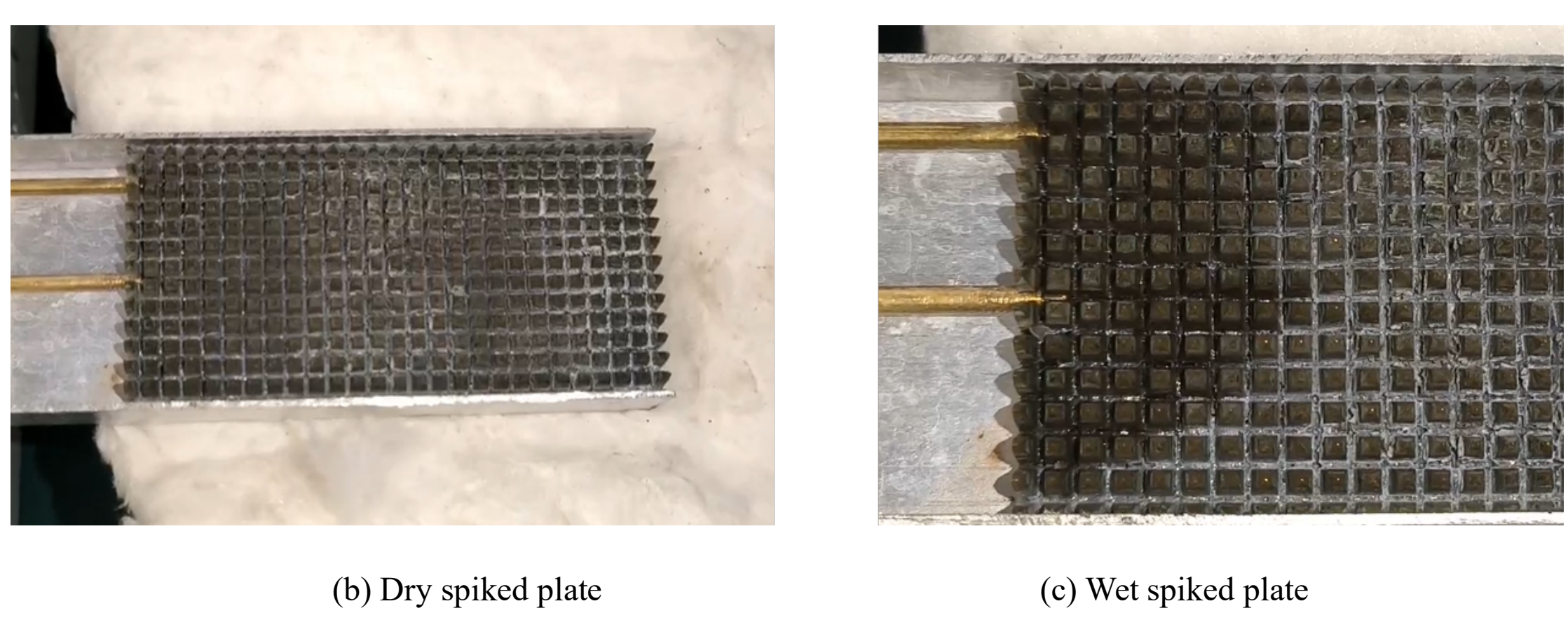

(b) Dry spiked plate

(c) Wet spiked plate

Fig. 7 Schematic Diagram of the Boiling Liquid Collection Device in Operation

## 2.4 Loop Heat Pipe Construction

Figure 8 shows a schematic diagram of the working principle of the new nanofluid heat pipe. The two-dimensional view can see the transportation process from the fluid and vapor heating zone to the condensing zone, and the three-dimensional view can see the circulation of the work material in the working process of the whole loop heat pipe. In the working process, the heating section uses the copper foam to strengthen the liquid boiling, the water vapor flows from the heating zone to the cooling zone, and condenses on the spiked plate; the splashed nanofluid flows into the collector and flows along the capillary tube to the spiked plate; the nanofluid flowing out of the capillary tube meets with the condensate, and is sucked in by the porous absorbent cotton on the left side, which flows back to the heating section and completes the cycle.

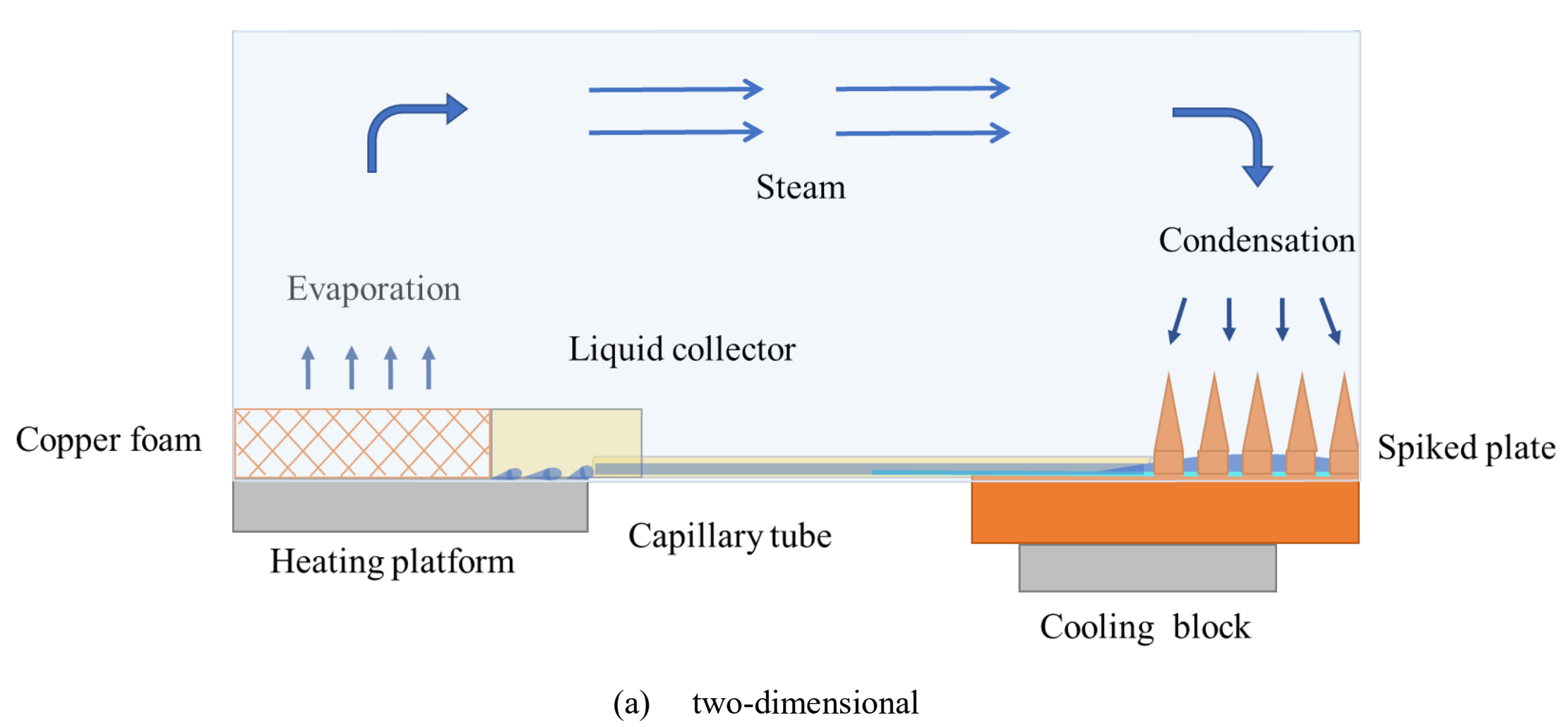

(a) two-dimensional

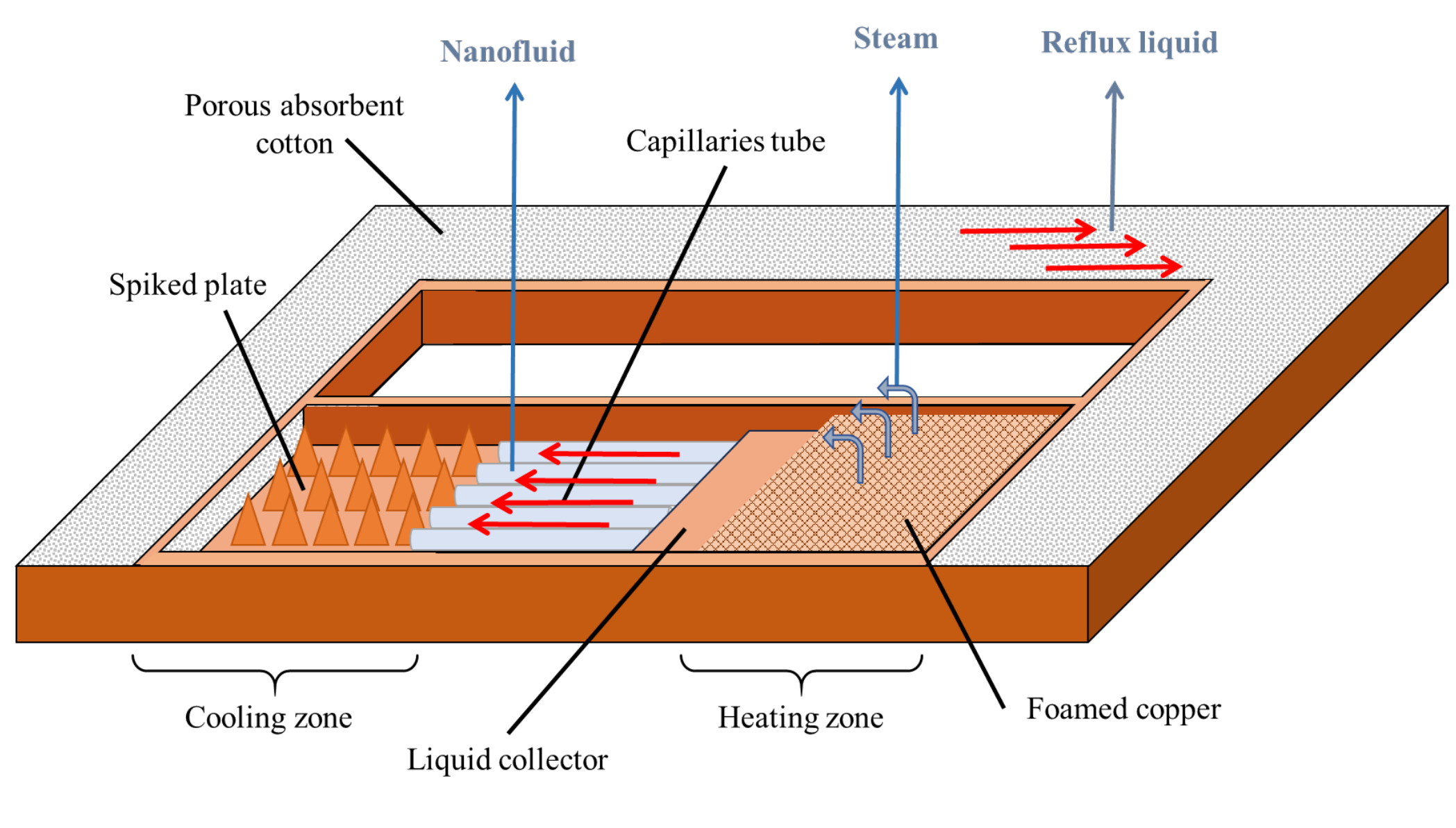

(b) Three-dimensional

Fig.8 Working principle diagram of nanofluid heat pipe.

## 2.5 Feasibility verification and gravity-independent verification

The visualized novel nanofluid heat pipe was made and running according to the schematic diagram, as shown in Figure 9. After the visible loop heat pipe worked stably for 40 minutes, the liquid was taken at the evaporating zone and condensing zone of the heat pipe respectively, by observing the color of the liquid as well as the nanoparticles left behind after drying to determine the difference in the concentration. Observed by eyes, there is no significant difference in the color of the work medium concentration in the evaporation and condensation zones as shown in the measuring cylinder. At the same time, pouring these fluids on the solid wall surface to make them evaporate naturally leaves the same traces. It can be seen that the nanoparticles are also involved in the whole cycle during the operation of the loop heat pipe. It proves the feasibility of the structure design.

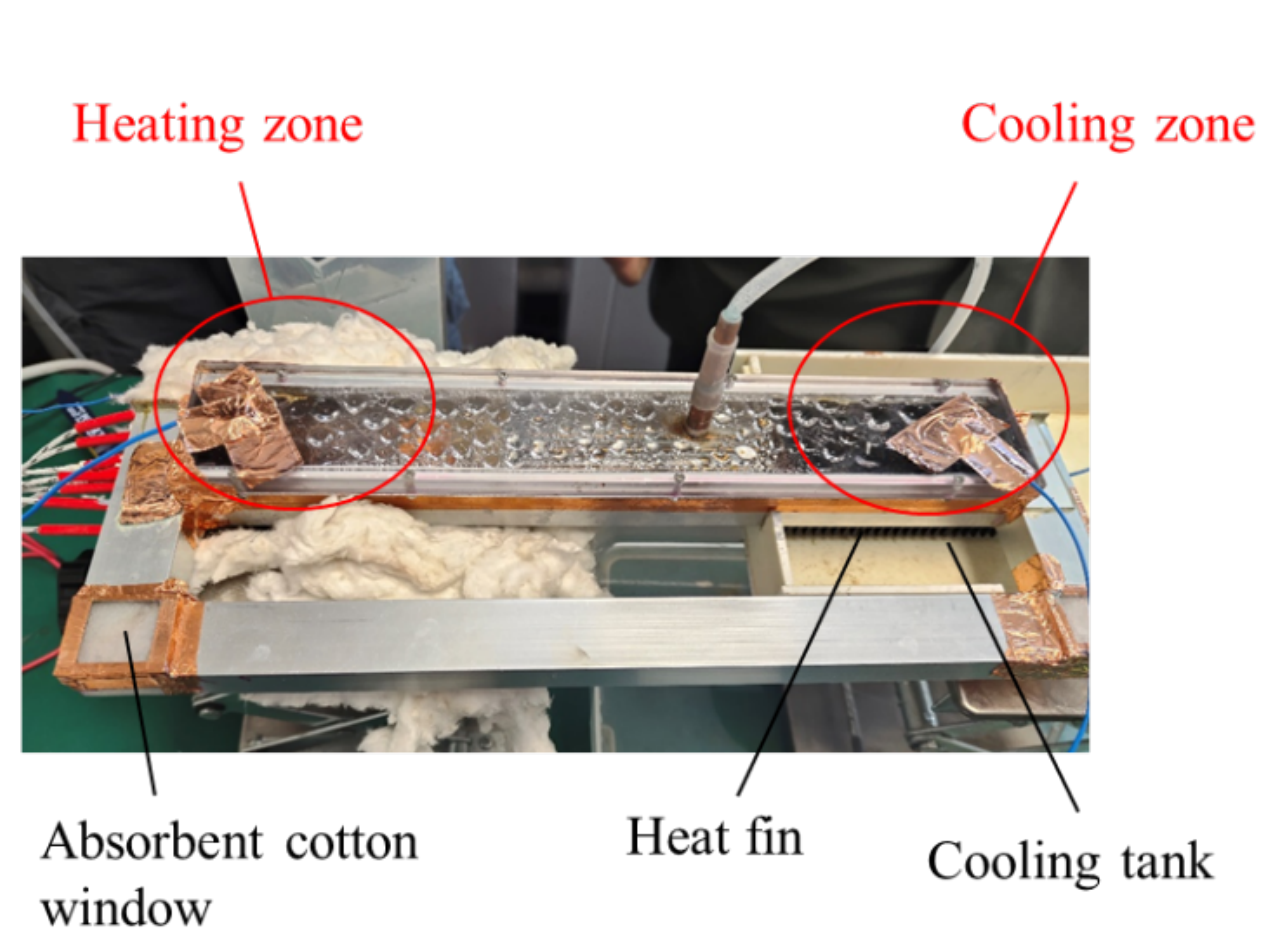

(a) Visible heat pipe work process

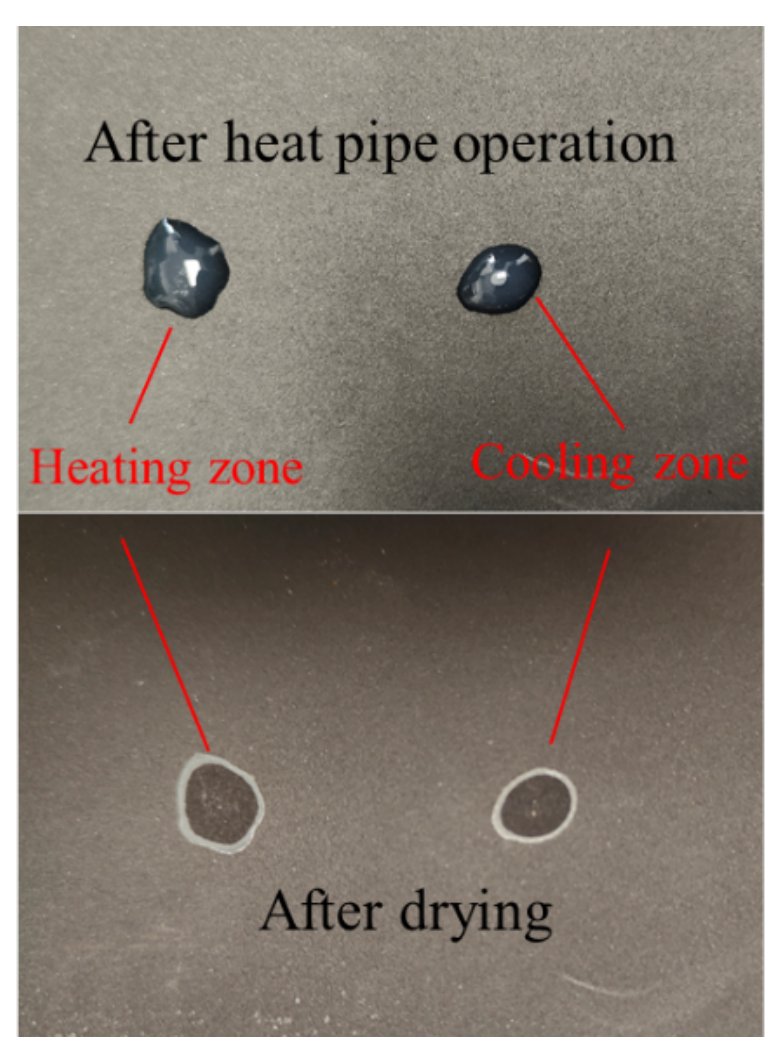

(b) Take the liquid after stabilizing the work

Fig.9 Nanofluid heat pipe normal operation

Design of gravity-independent verification experiments of the Nanofluid heat pipe on the ground. The heat pipe was placed horizontally normal (Figure 9a) and horizontally inverted (Figure 10), and the heat pipe worked normally in both cases. This shows that the vapor-liquid transport or phase transition mechanism in the heat pipe does not depend on gravity, thus the heat pipe can operate normally in the microgravity environment in space.

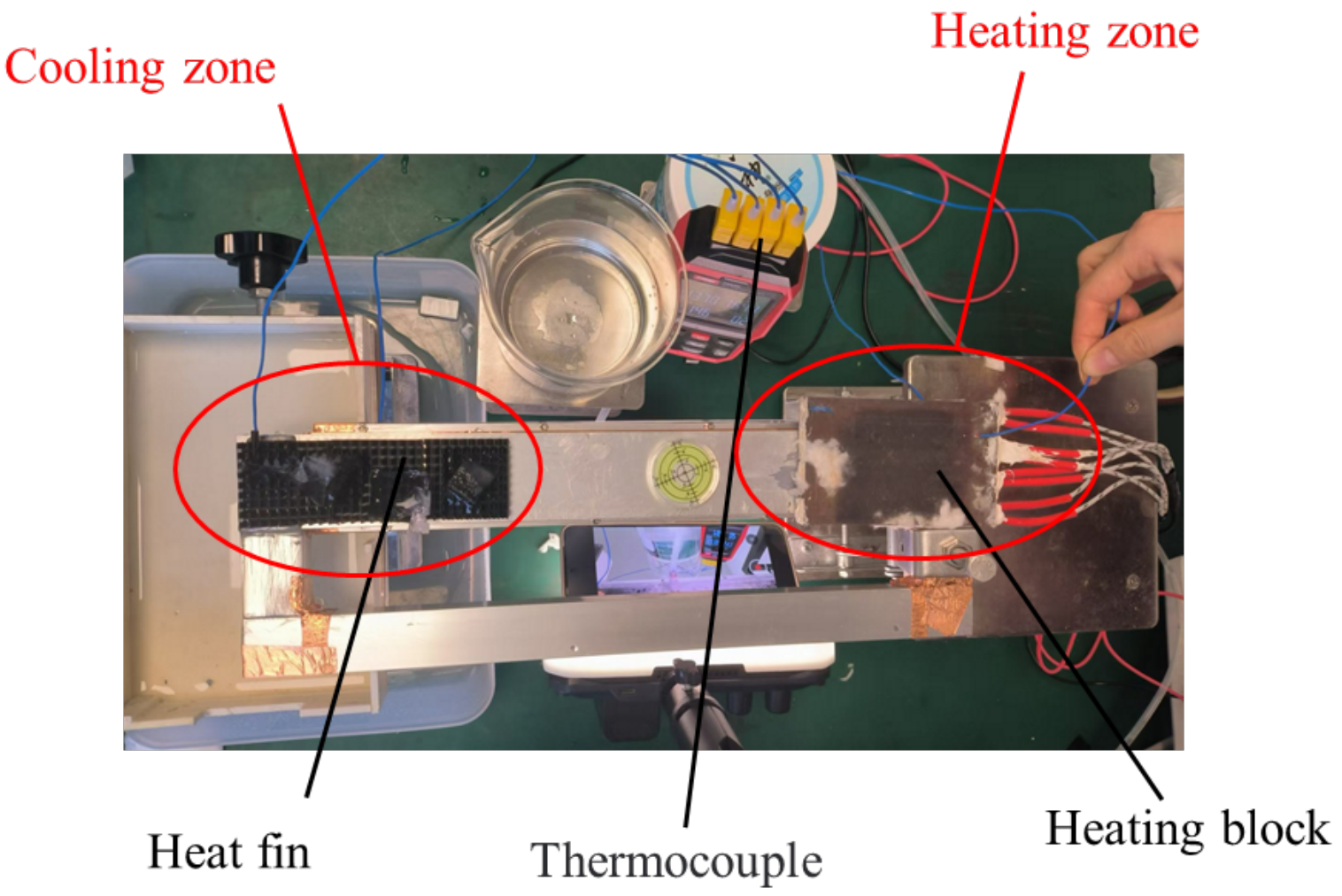

Fig.10 Gravity-independent verification of the working principle of nanofluidic heat pipes: Horizontal inversion experiment

Using the above technique, we tested the new technique of the whole convection of nanofluid (aqueous solution of alumina with 2% mass fraction) in the loop heat pipe. Nanofluids and vapors leave from heating zone and converge in cooling zone, then absorbed by porous water-washing cotton and redirected to the heating zone. This new nanofluid heat pipe achieves a stabilized power of 360 W. It is verified that the nanofluid has been convected throughout the loop heat pipe during the operation and it is independent of gravity.

## 3. Conclusions

This experiment investigated the phenomenon of enhanced boiling of water and nanofluids in copper foam on a heated plate, designed a novel nanofluid loop heat pipe, and reached the following conclusions:

(1) When liquid boils inside the copper foam, some of it is squeezed out along the heated surface. This phenomenon occurs with both water and nanofluids and is independent of gravity. From this, we can hypothesize that this phenomenon can also occur in a microgravity environment in space.

(2) The phenomenon of liquid being squeezed out of the copper foam during boiling is related to the hydrophilicity/hydrophobicity of the heating surface.

(3) The directional flow of the squeezed liquid is realized by structural design and a novel nanofluid loop heat pipe is designed. This solution solves the usual blockage problem of the capillary core of the nanofluid loop heat pipe.